 \documentclass[%
 prl,%
 amsmath,amssymb,
preprint,%
 reprint,%
twocolumn,
]{revtex4-2}
\usepackage{xcolor}
\usepackage{graphicx}
\usepackage{dcolumn}
\usepackage{bm}

\begin{document}

\preprint{AIP/123-QED}

\title{Unusually high-density 2D electron gases in N-polar AlGaN/GaN heterostructures with GaN/AlN superlattice back barriers grown on sapphire substrates}

\author{Maciej Matys}
 \altaffiliation{Fujitsu Limited, Atsugi, Kanagawa, 243-0197, Japan}
 \email{matys.maciej@fujitsu.com}
\author{Atsushi Yamada}
\affiliation{Fujitsu Limited, Atsugi, Kanagawa, 243-0197, Japan}
\author{Toshihiro Ohki}
\affiliation{Fujitsu Limited, Atsugi, Kanagawa, 243-0197, Japan}

\date{\today}

\begin{abstract}
We reported on the observation of extremely high-density ($>10^{14}$cm$^{-2}$) 2D electron gas in N-polar AlGaN/GaN heterostructures grown on sapphire substrates. Due to introducing the GaN/AlN superlattice (SL) back barrier between the GaN buffer layer and AlGaN barrier layer, we observed a giant enhancement of the 2D electron gas density at the GaN/AlGaN interface from $3\times10^{13}$cm$^{-2}$ (without SL) to $1.4\times10^{14}$cm$^{-2}$ (with SL back barrier) that is only one order of magnitude below the intrinsic crystal limit of $\approx10^{15}$cm$^{-2}$. We found that the changes of 2D electron gas density with SL correlated well with the changes of the wafer warp parameter which suggests that the strains are responsible for the 2D electron gas density enhancement (reduction of the piezoelectric polarization in the GaN channel). Nevertheless, this finding is probably insufficient to fully explain the observed high 2D electron gas density. Simultaneously, the room temperature electron mobility was 169 cm$^2$/Vs, which with the electron density of $1.4\times10^{14}$cm$^{-2}$ gives a low sheet resistance of 264 $\Omega$/sq (one of the lowest reported so far for the N-polar 2D electron gas channel). Finally, the possibility of application of such high-density 2D electron gas with low sheet resistance to transistors, emitters and detectors was discussed.
\end{abstract}
\maketitle

The discovery of the high-conductivity quantum-confined two-dimensional (2D) electron gases at the interface of AlGaN/GaN semiconductor heterostructures in the mid-1990s\cite{K1} has generated considerable attention in the scientific community since its formation did not require the presence of chemical dopants. Contrary to AlGaAs/GaAs material systems\cite{H1,H2,H3}, 2D electron gases in AlGaN/GaN heterostructures are formed spontaneously due to net polarization charges composed of spontaneous and piezoelectric polarization of AlGaN and GaN\cite{A,b}.

\begin{figure}
\includegraphics{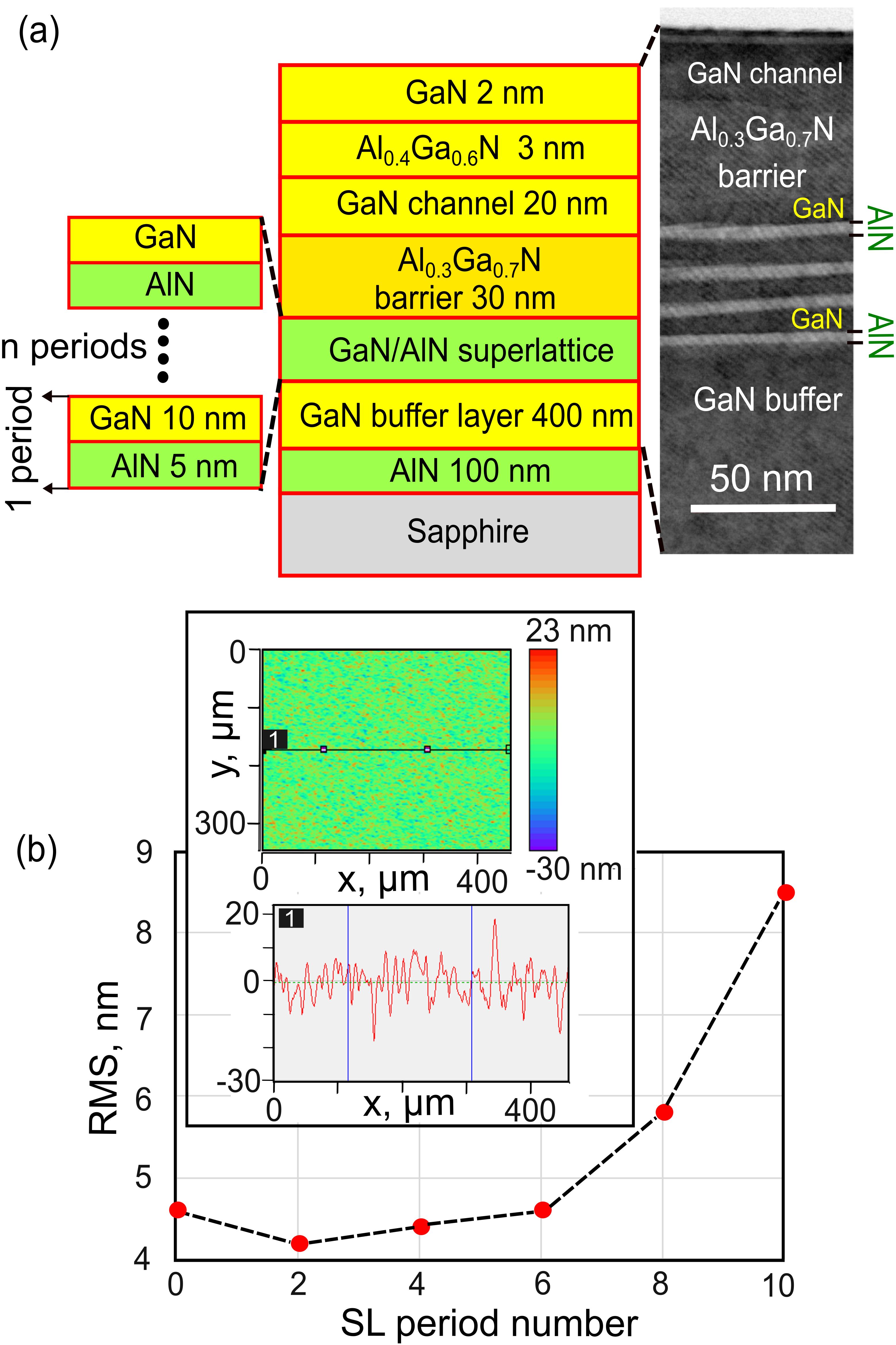}
\caption{\label{fig:epsart} (a) Schematic illustration of the fabricated epitaxial structures used in this study. (b) Root mean square (RMS) roughness as a function of SL periods. Inset of (a) shows the TEM image of the AlGaN/GaN heterostructure with 4 period GaN/AlN SL structure. Inset of (b) shows the 2D and 1D surface profiles from interferometric surface profilers of the AlGaN/GaN heterostructure with 4 period GaN/AlN SL}
\end{figure}

Recently, the N-polar III-nitride heterostructures attracted great attention in nitride community due to their several advantages over metal-polar counterpart in high frequency and high-power applications\cite{r1,b1,b2,b3,b4,s1,s2}. The typical as-grown 2D electron gas density in N-polar III-nitride heterostructures ranging from low-$5\times10^{12}$ to high $3\times10^{13}$cm$^{-2}$ with mobility between 300–2500 cm$^2$/Vs and sheet resistances from 150 to 600 $\Omega$/sq\cite{r1,z1,z2,K1,K2,K3}. However, it should be highlighted the highest observed 2D electron gas densities in N-polar III-nitride heterostructures were $4.3\times10^{13}$cm$^{-2}$ with mobility 450 cm$^2$/Vs and sheet resistance 320 $\Omega$/sq\cite{z1,z2}. Such gases were realized using a high Al-content AlGaN barrier and AlN substrates\cite{z1,z2}. The 2D electron densities higher than $10^{14}$cm$^{-2}$ have not been measured in N-polar III-nitride semiconductor heterostructures so far in an as-grown structure or field-effect induced channel.

	In this work, the extremely high-2D electron densities ($>10^{14}$cm$^{-2}$) are demonstrated in N-polar AlGaN/GaN heterostructure. It was found that introducing the GaN/AlN superlattice (SL) back barrier between GaN buffer layer and AlGaN barrier layer leads to a giant enhancement of 2D electron gas density at GaN/AlGaN interface. In particular, we observed an increase of 2D electron gas density from $3\times10^{13}$cm$^{-2}$ (without SL structure) to $1.4\times10^{14}$cm$^{-2}$ (with SL) that is only an order of magnitude below the intrinsic crystal limit of $\approx10^{15}$cm$^{-2}$. The Hall measurement shows that this high-2D electron gas density has a room temperature electron mobility as 169 cm$^2$/Vs which with 2D electron gas density of $1.4\times10^{14}$cm$^{-2}$ leads to a very low sheet resistance of 264 $\Omega$/sq.

\begin{figure}
\includegraphics{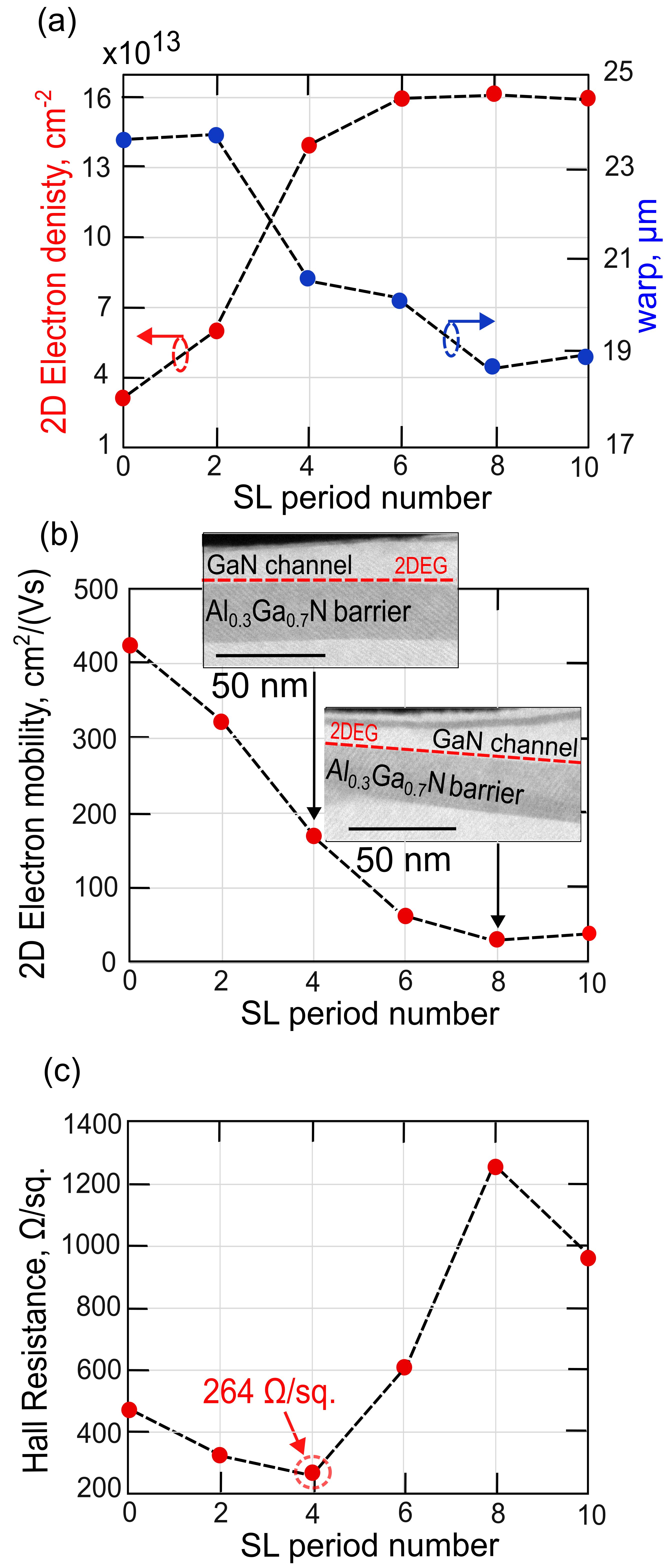}
\caption{\label{fig:epsart} (a) Room-temperature 2D electron gas density and warp paramter as a function of the SL period number, (b) room-temperature electron mobility vs. SL period number and (c) room-temperature Hall sheet resistance as a function of the SL period number. Inset of (b) shows TEM images of the GaN channel region of structures with (a) 4 and (b) 8 period SL.}
\end{figure}

Fig. 1(a) shows the schematic illustration of the fabricated N-polar AlGaN/GaN heterostructures. The investigated structures were grown by Metal-Organic Chemical Vapour Deposition (MOCVD) using trimethylgallium (TMGa), trimethylaluminum (TMAl), and ammonia as precursors on (0001) sapphire with a misorientation angle of 4$^o$ toward the a-sapphire-plane. Firstly, a 100 nm AlN layer was grown at  1025 $^o$C with a rate of 2 $\mu$m/h. Subsequently, a 400 nm unintentionally doped GaN buffer layer was grown at 1025 $^o$C with a V/III ratio of 1600 at the pressure of 20 kPa. On top of the buffer GaN layer a thin superlattice (SL) structure was deposited, which contains the alternating GaN and AlN layers (see inset of Fig. 1(a)). The thickness of GaN and AlN layers in SL structure was 10 and 5 nm, respectively. The number of SL periods, n, (inset of Fig. 1(a)) was varying between 0 to 10. Next, a 30-nm-thick Al$_{0.3}$Ga$_{0.7}$N barrier layer was deposited followed by a 20 nm thick undoped GaN channel layer. Finally, 3-nm-thick Al$_{0.4}$Ga$_{0.6}$N and 2-nm-thick GaN cap layers were grown on the top of the structure. The polarity of the fabricated N-polar AlGaN/GaN heterostructures was confirmed by the KOH etching and X-ray diffraction (XRD) measurements (not shown). The transmission electron microscope (TEM) image confirmed that the AlN/GaN SL structure was successfully grown (see inset of Fig. 1(a)). The surface morphologies of the fabricated N-polar AlGaN/GaN heterostructures were measured using interferometric surface profiler. The  measurement from the large area over 400 $\mu$m 300 $\mu$m indicates the smooth morphology with a root mean square ($RMS$) roughness between 4.2 to 4.5 nm for 0 to 6 period SL as shown in Fig. 1(b). However above 6 period SL, the surface becomes very rough with $RMS$ value of 8.5 nm for 10 period SL (Fig. 1(b)).

\begin{figure}
\includegraphics{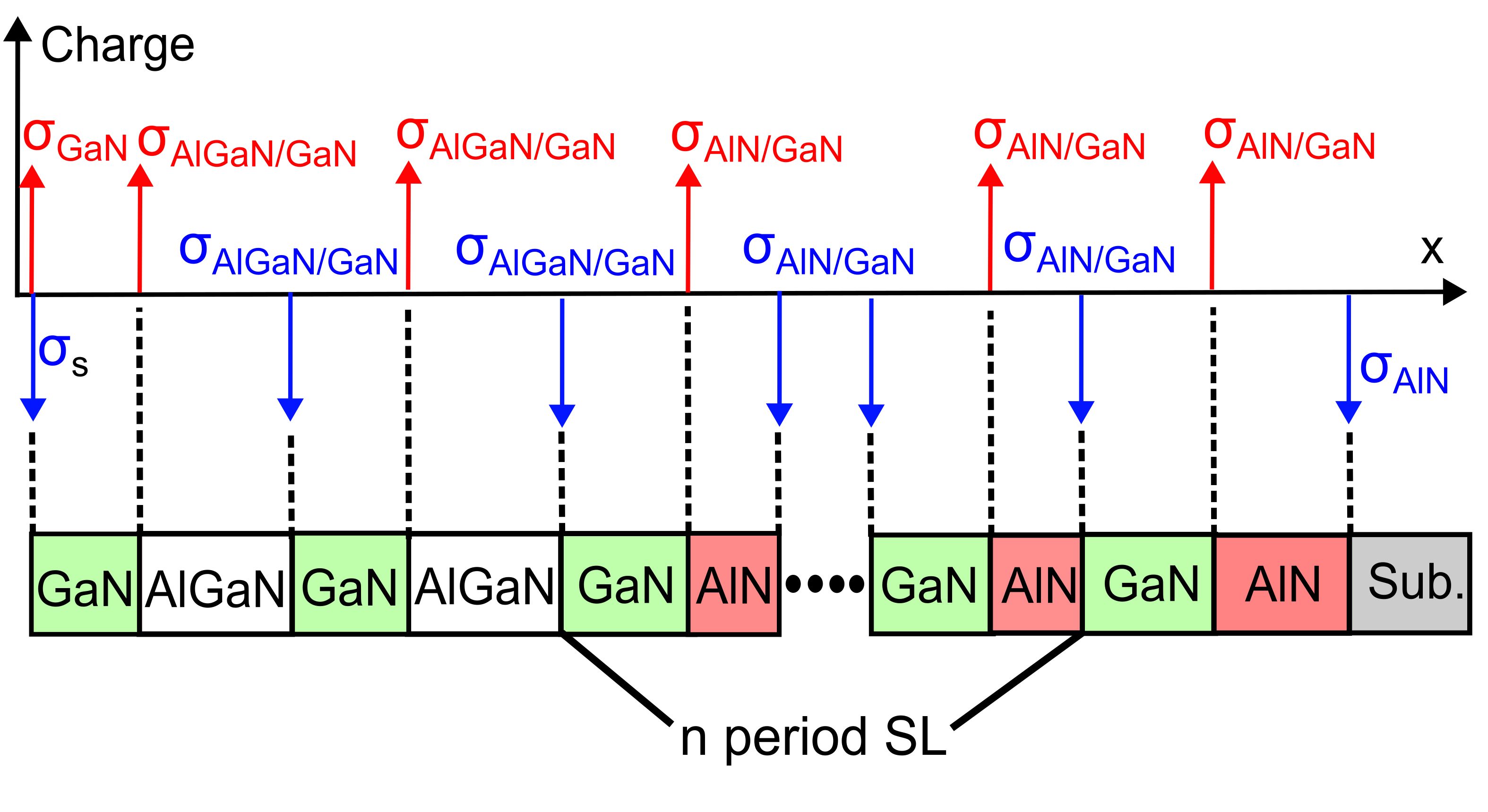}
\caption{\label{fig:epsart} Distribution of polarization charges in the investigated SL based N-polar AlGaN/GaN heterostructures. }
\end{figure}
\begin{figure}
\includegraphics{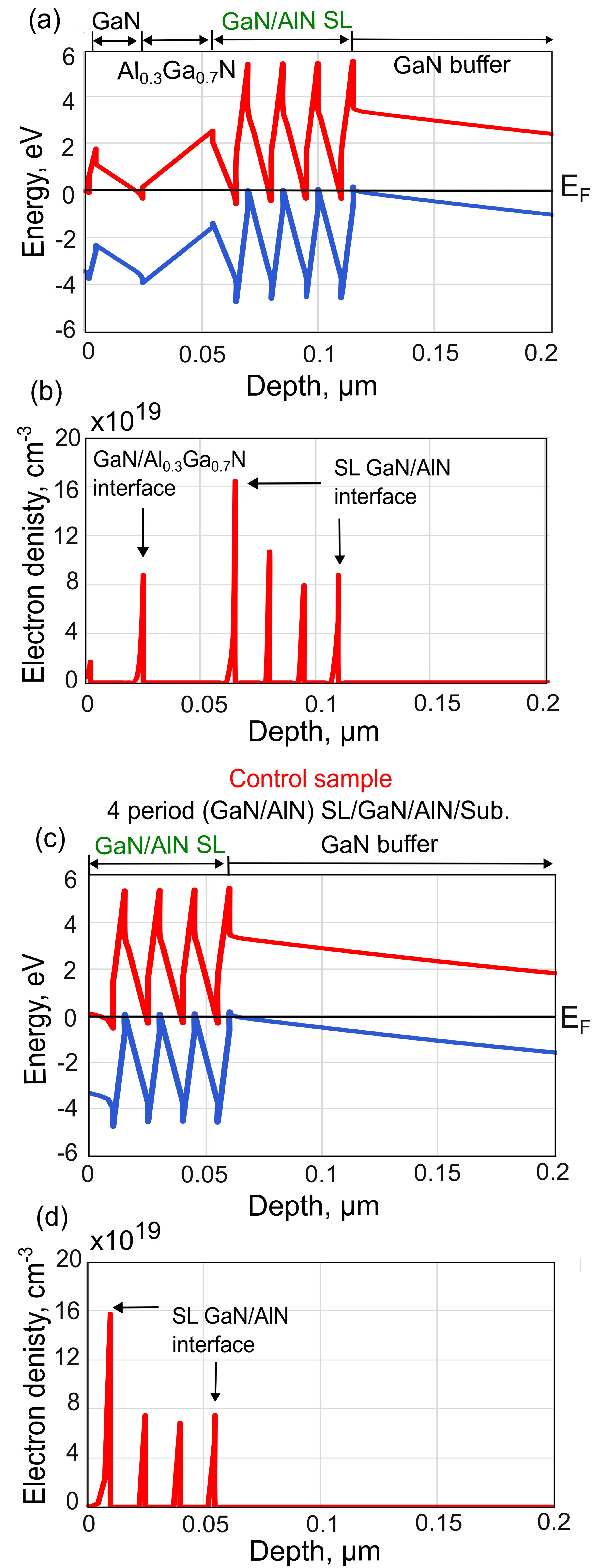}
\caption{\label{fig:epsart} Simulated (a) band diagram and (b) electron density of AlGaN/GaN heterostructure with 4 period GaN/AlN SL structure. Simulated (c) band diagram and (d) electron density of the control sample, i.e., 4-period (GaN/AlN) SL/GaN (400 nm)/AlN (100 nm)/sapphire.}
\end{figure}

To characterize the electron transport properties in the fabricated N-polar AlGaN/GaN heterostructures, we applied the Hall-effect measurements with the Van der Pauw method using the aluminum-titanium Ohmic contacts and the magnetic field of 1 T. Fig. 2(a) and (b) show the measured room temperature 2D electron density and mobility as a function of the SL period number. Without SL structure the N-polar AlGaN/GaN heterostructures exhibits the 2D electron gas density of $3\times10^{13}$cm$^{-2}$ with mobility of 440 cm$^2$/Vs. The mobility value of 440 cm$^2$/Vs is relatively low due to the non-uniformity of AlN/GaN SL (see the TEM image, inset of Fig. 1(a)) which arises as a result of the growth on the sapphire substrate with a miscut angle orientation. By increasing the SL period number, the 2D electron gas density initially slowly increases and then jumps rapidly at 4 period SL reaching the concentration of $1.4\times10^{13}$cm$^{-2}$ (see Fig. 2(a)). Further, the increasing of SL periods leads to the saturation of the electron gas density at the level $1.6\times10^{13}$cm$^{-2}$. Simultaneously, the electron mobility decreases with increasing of SL periods up to 8 and then becomes constant (Fig. 2(b)). We attributed this effect to the interface roughness scattering mechanism. In particular, Singisetti et al.\cite{K2} showed that the interface roughness is a dominant scattering mechanism limiting the electron mobility in the N-polar GaN quantum well channels. On the other hand, the interface roughness scattering is proportional to the square of the electric field, which likely increases in the region of GaN/AlGaN interface with the SL period number increase. However, this is only the speculation and thus in order to clarify the cause of mobility reduction, we performed a more detailed analysis of the interface roughness using TEM. The inset of Fig. 2(b) shows TEM images of the GaN channel region of structures with 4 and 8 period SL. As can be seen, in the case of the structure with 4 period SL, the GaN/Al$_{0.3}$Ga$_{0.7}$N interface (where 2DEG is located) is relatively flat while in the structure with 8 period SL the GaN/Al$_{0.3}$Ga$_{0.7}$N interface is non-flat, i.e. strongly tilted downwards. Because this tilted one appears already at the short distance of 50 nm, it is very likely that the GaN/Al$_{0.3}$Ga$_{0.7}$N interface of the structure with 8 period SL is very rough. This conclusion is also well consistent with RMS analysis using interferometric surface profilers (see Fig. 1 (b)). Thus, enhancement of the interface roughness is a reasonable cause of mobility reduction with the SL period number increase. Fig. 2(c) shows the room temperature Hall sheet resistance ($R_H$) as a function of the SL period number. As can be seen, $R_H$ exhibits a U-shaped dependency on the SL period number, namely with increasing n, $R_H$ gradually decreases reaching the minimum at 4 period SL and subsequently increases. In the range from 0 to 4 period SL, RH decreases almost 2 times which is a direct consequence of the rapid increasing of the 2D electron gas density (Fig. 2(a)). On the other hand, in the range from 4 to 8 period SL, RH increases approximately 4.5 times. This increases is due to the fact that the 2D electron density is practically saturated in the range from 4 to 8 period SL. However, the electron mobility in this range still continues decreasing from 169 at 4 period SL to 80 cm$^2$/Vs at 8 period SL (Fig. 2(b)).

To explain the above results, we performed the Technology Computer Aided Design (TCAD) simulations of the band diagram of investigated heterostructures. In the simulation, we assumed the polarization charges ($\sigma$) at AlN/GaN, Al$_{0.3}$Ga$_{0.7}$N/GaN and Al$_{0.4}$Ga$_{0.6}$N/GaN interfaces given by the following equation\cite{A}: 

\begin{equation}
\frac{\sigma}{q}=6.41\times10^{13}x-1.17\times10^{13}x(1-x)
\end{equation}

where $\sigma$ is in the unit of cm$^{-2}$ and $x$ is the Al composition. In addition, we assumed that on the GaN surface the positive spontaneous polarization charge of $\sigma_{GaN}$=$1.8\times10^{13}$cm$^{-2}$ is compensated by the surface negative charge $\sigma_{S}$=-$1.8\times10^{13}$cm$^{-2}$. Such compensation charge was previously found on the N-polar GaN surface by the X-ray photoelectron spectroscopy measurements\cite{E1}. The detailed distribution of the assumed polarization charges in the heterostructure is shown in Fig. 3.

The band diagram simulation clearly shows that in addition to the 2D electron channel at the GaN/AlGaN interface, the high-density 2D electron gases should be formed in the SL structure (see Figs. 4(a) and (b)). The formation of the parallel 2D electron channels in the SL structure could explain a significant increase of the 2D electron gas density between 0 and 4 period SL (Fig. 2(a)). However, the saturation of 2D electron gas density between 4 to 10 period SL is the signature that 2D electron gas is not formed in the SL structure (2D electron gas density should depends on the SL periods number approximately linearly, as it was reported previously). To confirm no 2D electron gases in the SL structure we growth the control sample i.e. 4-period SL/GaN (400 nm)/AlN (100 nm) /Sapphire. The band diagram simulations indicated that the SL structure in the control sample should contain the similar very high denity-2D electron gas as investigated ones (see Figs. 4(c) and (d)). However, the Hall-effect measurements from the control sample shows high contactless sheet resistance of 710 Ohm/s, which is similar to the sheet resistance of the 400 nm N-polar GaN layer (800 Ohm/sq). This means that 2D electron gas was not formed in the SL structure. One of the reasons why the 2D electron gas in the SL structure is not observed may be the fact that AlN in the SL is not pure AlN but a high content AlGaN. For example, if instead of the AlN layer the Al$_{0.6}$Ga$_{0.4}$N layer has been formed in SL the 2D electron gas should not appear, as shown in Figs. 5(a) and (b). Another reason of high resistivity of the SL structure can be existence of acceptor traps at the negative polarization interfaces of SL\cite{r1}.

\begin{figure}
\includegraphics{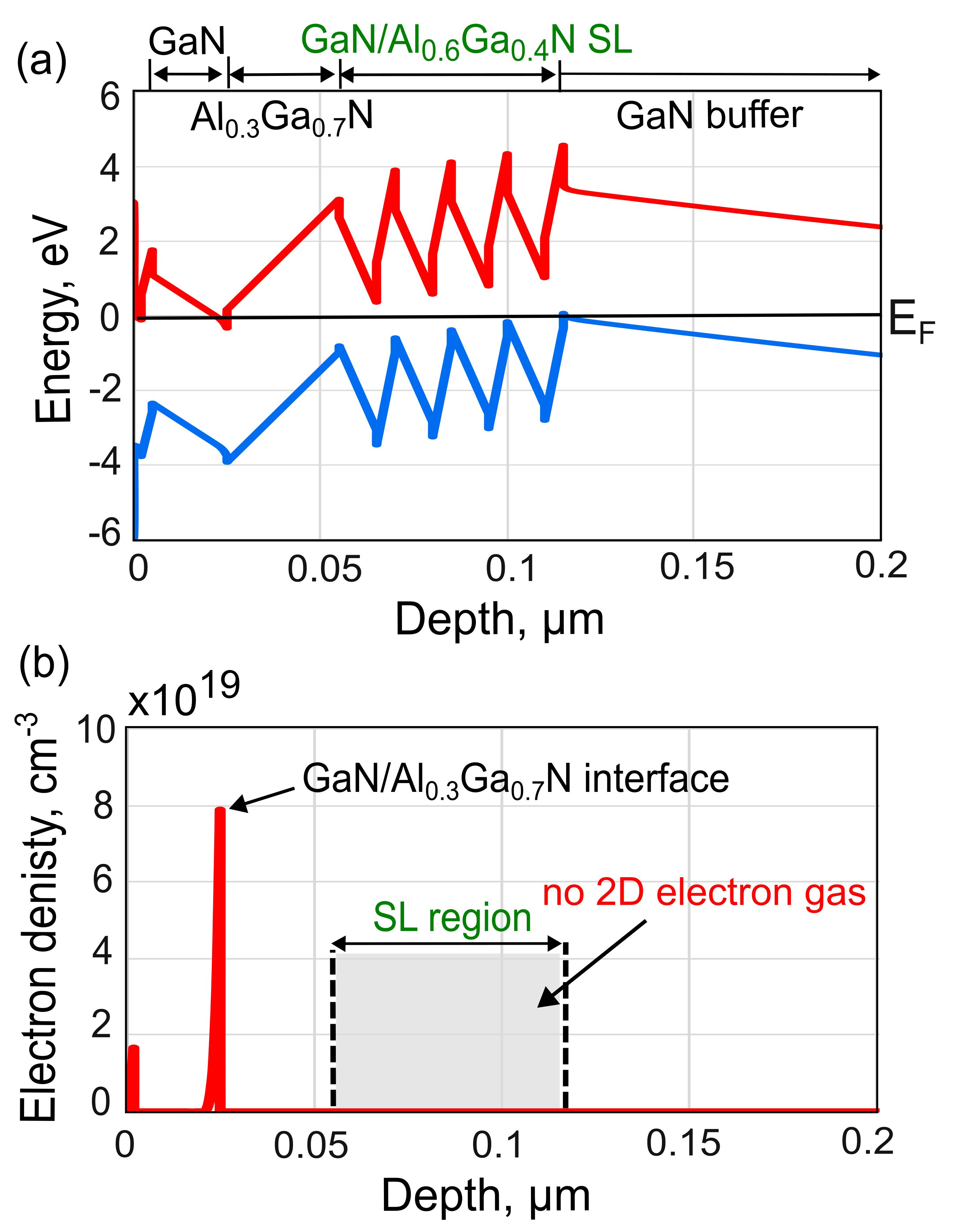}
\caption{\label{fig:epsart} Simulated (a) band diagram and (b) electron density of AlGaN/GaN heterostructure with 4 period GaN/Al$_{0.6}$Ga$_{0.4}$N SL structure.}
\end{figure}

\begin{figure}
\includegraphics{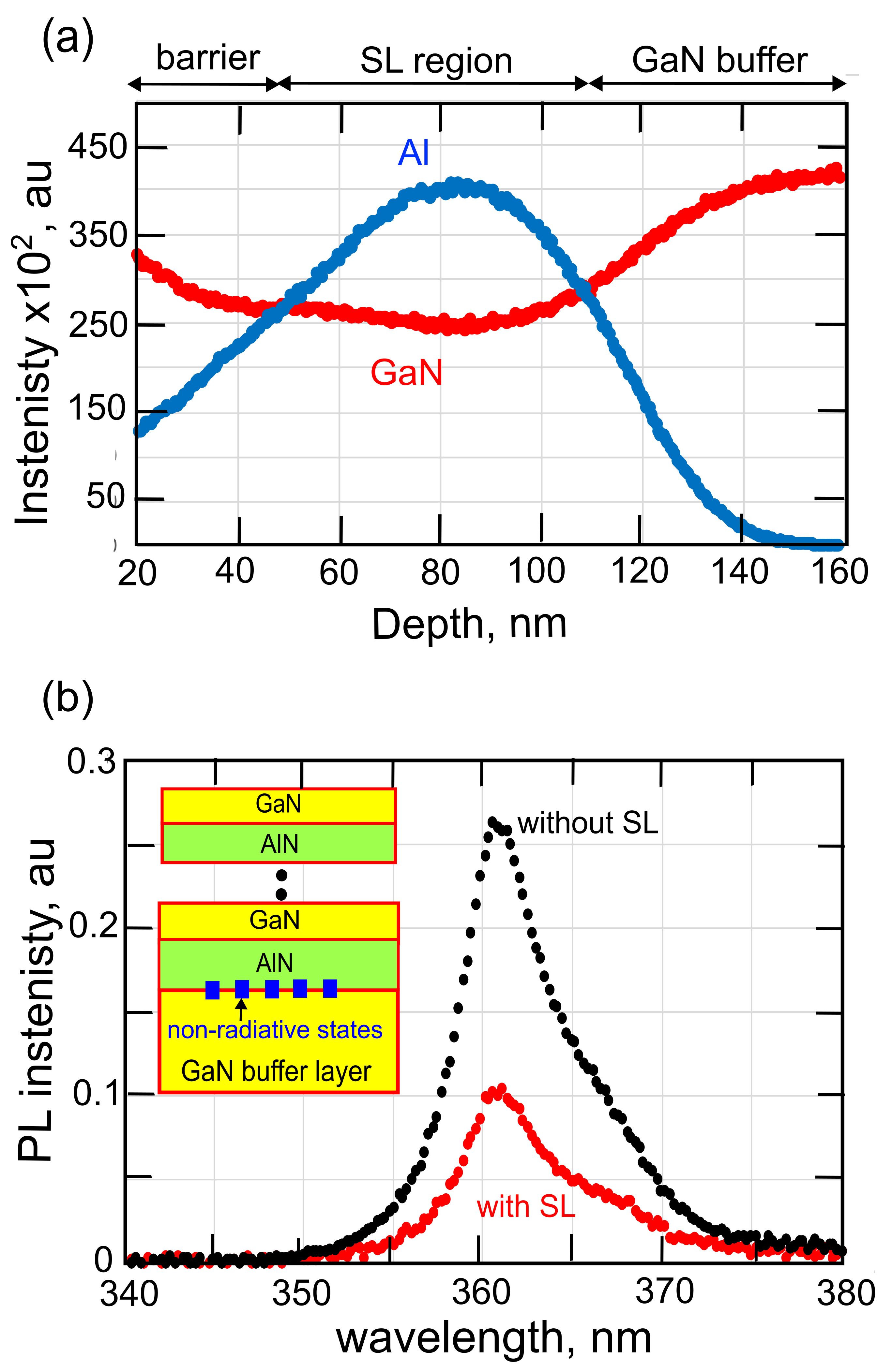}
\caption{\label{fig:epsart} (a) SIMS depth profile of the control sample, i.e. 4-period (GaN/AlN) SL/GaN (400 nm)/AlN (100 nm)/sapphire. (b) PL spectra in the 340-380 nm range from the samples without SL and 4 period SL.} 
\end{figure}

To verify which hypothesis is more likely, we performed chemical characterization of the control sample (4-period (GaN/AlN) SL/GaN (400 nm)/AlN (100 nm)/sapphire) using Secondary Ion Mass Spectrometry (SIMS), as shown in Fig. 6(a). One can note from this figure that Ga atoms are present in the whole SL region at a similar content level which indicates that the AlN layers in SL are not pure AlN but rather high Al-content AlGaN (in the case of AlN layers, we should observe some periodicity of Ga in the SL region). In addition, to identify the presence of traps at the SL interfaces, we performed the photoluminescence (PL) measurements of structures with and without SL. Fig. 6(b) shows the PL spectra in the 340-380 nm range from the samples without SL and 4 period SL. The PL peak which is related mainly to the band-to-band recombination in the GaN buffer layer is clearly lower in the case of structures with SL (about 3 times compared to the structure without SL). The reduction of band-edge PL peak can be attributed to the presence of non-radiative traps at the AlN (SL)/GaN (buffer layer) interface, as schematically shown in the inset of Fig. 6(b). More precisely, these traps reduce the band-to band recombination in the GaN buffer layer by the non-radiative interface recombination. If such traps are present at the AlN (SL)/GaN (buffer layer) interface, we can assume that they are also present at AlN/GaN SL interfaces. Therefore, all these results means that we cannot clearly state, which hypothesis is true.

In light of the lack of a 2D electron gases in SL structure we can assume that the carrier transport in the investigated heterostructures (Fig. 1(a)) was mainly by the 2D electron gas at the GaN/Al$_{0.3}$Ga$_{0.7}$N interface. However it should be noted that besides 2D electron channel at the GaN/Al$_{0.3}$Ga$_{0.7}$N interface there is also some 2D electron channel in GaN cap layer (see Figs. 5(a) and (b)). According to the TCAD simulation this 2D electron gas has totaly negligible concentration compare to the 2D electron gas at the GaN/Al$_{0.3}$Ga$_{0.7}$N interface (Fig. 5(b)). Thus it can be conclude that carrier transport was entirely dominant by the 2D electron gas at the GaN/Al$_{0.3}$Ga$_{0.7}$N interface.

\begin{figure}
\includegraphics{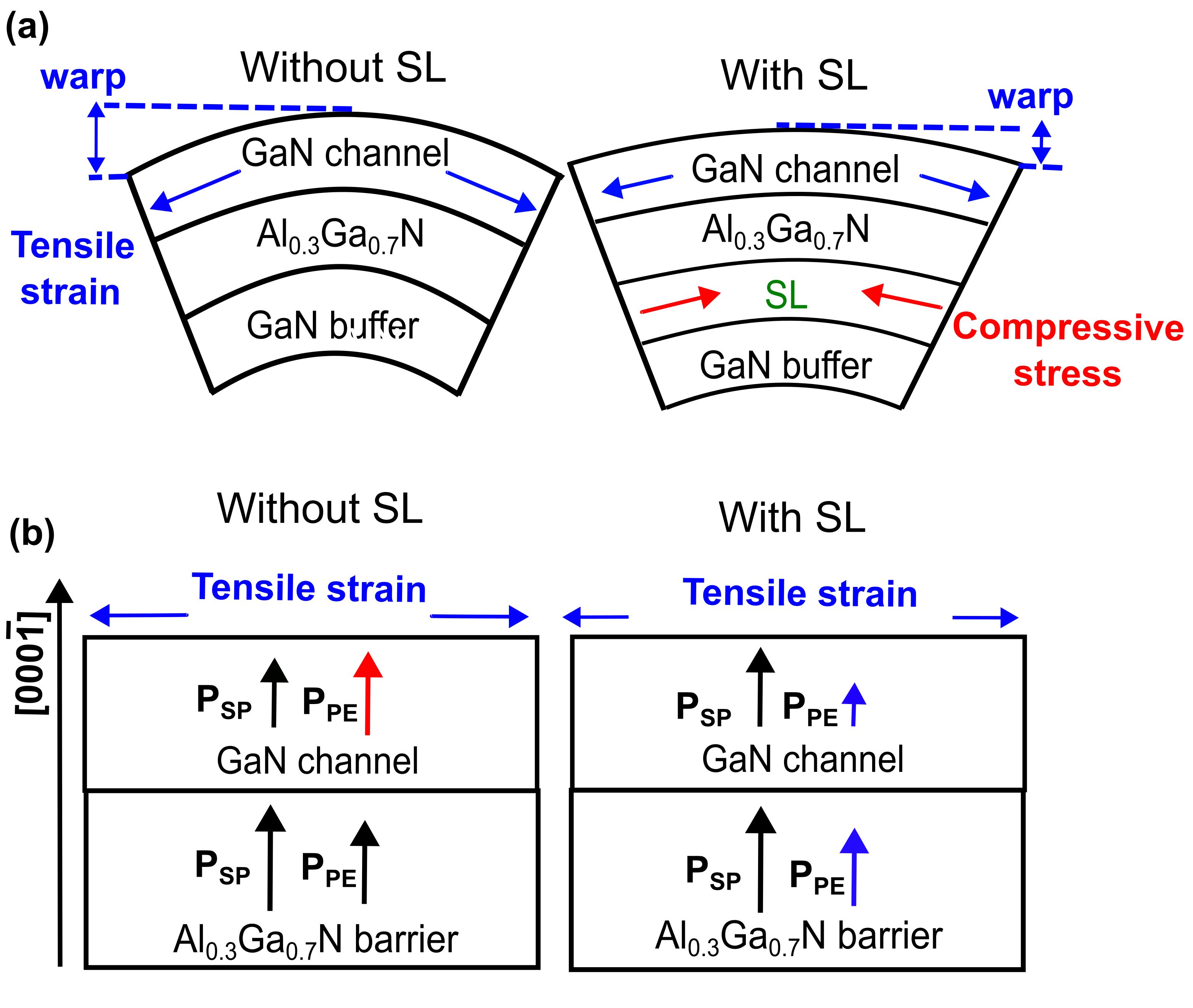}
\caption{\label{fig:epsart} (a) Schematic illustration of warp parameter reduction in the investigated epitaxial structures. (b) Directions of the spontaneous and piezoelectric polarizations in the examined structures. Enhancement of 2D electron density at the GaN/Al$_{0.3}$Ga$_{0.7}$N interface occurred due to reduction of $P_{PE}$ in GaN channel after SL deposition.}
\end{figure}

After excluding the formation of 2D electron gas in the SL structure another reason for enhancement of 2D electron gas density needs to be identify. The enhancement of the stress due to SL deposition is a natural factor that could cause increasing of the 2D electron gas. For example, previously K.-S. Im et al.\cite{KS} reported a giant formation of 2D electron gas density $>10^{14}$cm$^{-2}$ in the metal-polar AlGaN/GaN heterostructure on Si substrates by generation of a strong tensile stress. To estimate the strain in the fabricated heterostructures, we measured the standardized warp parameter (flatness of the wafer) as shown in Fig. 2(a). The relationship between the warp parameter ($w$) and residual stress in the film ($\sigma$) can obtained from the Stoney’s formula:\cite{Ost,Ost2}

\begin{equation}\label{wz_2}
   \sigma_f=\frac{E_sh_s^2}{6Rh_f(1-v_s)}
\end{equation}

where $R$ is the radius of plate curvature, $E$ is the Young’s modulus, $h$ is the thickness, $v$ is the Poisson's ratio, and the subscripts f and s correspond to the film and substrate, respectively\cite{Ost,Ost2}. Eq. 2 can be applied when the film is much thinner than the substrate and radius of curvature is much larger than the substrate thickness ($R\gg h_s\gg h_f$). In our case these conditions are fully satisfied. On the other hand from the geometrical considerations, the radius of curvature can be expressed by the warp parameter as:\cite{Ost,Ost2} 

\begin{equation}\label{wz_2}
   w\approx \frac{d^2}{8R}
\end{equation}

where $d$ is the wafer diameter. When we combine Eqs. 2 and 3, we obtain the following expression for $\sigma$ in terms of $w$:

\begin{equation}\label{wz_2}
   \sigma_f=\frac{8wE_sh_s^2}{6d^2h_f(1-v_s)}
\end{equation}

The above equation shows that the changes of the warp parameter directly reflect changes of the residual stress in the film. On the other hand, as can be seen from Fig. 2(a), the dependencies of the warp parameter as a function of the SL period number is strikingly similar to the dependence of 2D electron gas vs. SL period number. In particular, the biggest changes of the warp parameter occur between 2 and 4 period SL, where the giant enhancement of 2D electron gas density takes place (Fig. 2(a)). Furthermore, above 4 period SL, it seems that the warp parameter starts to saturate similarly like 2DEG density. Therefore, according to Eq. 4, enhancement of 2DEG density between 2 and 4 period SL and its subsequent saturation may directly results from the changes of residual stress in the film. A possible mechanism can be as follows. Decreasing of the warp parameter in the range between 2 and 4 period SL indicates that the wafer becomes more flatten. On the other hand the bow parameter was positive which means that the curvature of the wafer was concave. Thus the wafer flattening occurred probably due to the generation of compressive stress in SL structure which reduced tensile strain as schematically shown in Fig. 7(a). The reduction of tensile strains can leads to enhancement of the 2D electron density at the GaN/Al$_{0.3}$Ga$_{0.7}$N interface in the following manner. Firstly we recall the definition of the piezoelectric polarization $P_{PE}$ vector\cite{Mar}

\begin{equation}
P_{PE}=2\epsilon_{xx}(\frac{e_{31}-e_{33}C_{13}}{C_{33}})
\end{equation}

where $e_{31}$ and $e_{33}$ are the piezoelectric coefficients, $\epsilon_{xx}$ is the in-planecomponent of the strain tensor, and $C_{13}$ and $C_{33}$ are components of the compliance tensor. 

Form above equation it follows that if the GaN layer is under the tensile strain, $P_{PE}$ vector in GaN channel has the same direction as spontaneous polarization ones $P_{SP}$ (see Fig. 7(b)) since the factor $\epsilon_{xx}(\frac{e_{31}-e_{33}C_{13}}{C_{33}})$$>$0. This means that $P_{PE}$ according to Fig. 7(b) decreases 2D electron gas density at the GaN/Al$_{0.3}$Ga$_{0.7}$N interface. On the other hand as we mentioned before the tensile strain become reduced when SL is deposited (Fig. 7(a)). Thus in the AlGaN/GaN heterostrucutres with SL, $P_{PE}$ in GaN channel should be lower than in the structures without SL (see Fig. 7(b)). If we assumed now that decreases of $P_{PE}$ in Al$_{0.3}$Ga$_{0.7}$N barrier layer after deposited of SL is much lower in the case of GaN channel (Fig. 6 (b)) then enhancement of 2D electron density at the GaN/Al$_{0.3}$Ga$_{0.7}$N interface will occur due to reduction of $P_{PE}$ in GaN channel. However there is one weak point of this explanation: changes of $P_{PE}$ in GaN channel itself may be too weak to induced such huge 2D electron density (Fig. 2(a)). More precisely, $P_{PE}$ in GaN depends on $\varepsilon_{xx}$ according to the following formula:\cite{Mar} $P_{PE}(C/cm^{2})=-1.5\times10^{-6}\varepsilon_{xx}$. On the other hand, the typical range of $\varepsilon_{xx}$  is $<$2$\%$. This means that even if we assume that in the structure without SL the in-plane tensile strain is as high as 2$\%$, we should observe 2DEG density increase by $\approx2\times10^{13}$cm$^{-2}$  with SL due to reduction of the in-plane tensile strain and $P_{PE}$ in the GaN layer, which is clearly too low. Thus we believe that there are other stress-related contributions leading to enhancement of the 2D electron density at GaN/Al$_{0.3}$Ga$_{0.7}$N interface. Another factor which could enhancement the 2D electron density is that introduced SL may act as excellent electron blocking layers suppressing the escape of electrons from quantum well at GaN/Al$_{0.3}$Ga$_{0.7}$N interface.

\begin{figure}
\includegraphics{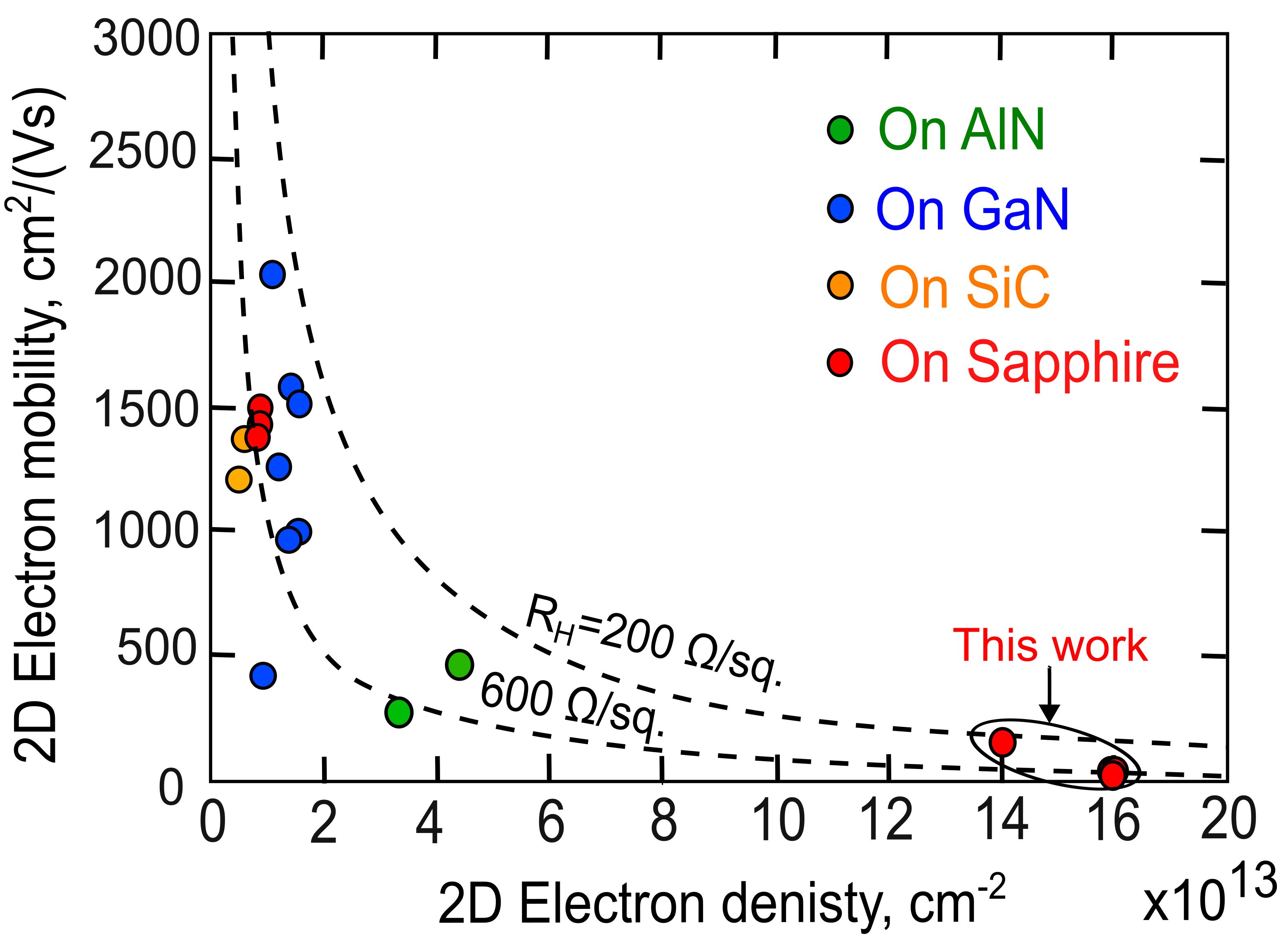}
\caption{\label{fig:epsart} Room-temperature electron density and mobility reported in N polar III-nitride semiconductor heterostructures grown on various substrates: sapphire, GaN and AlN\cite{r1,z1,z2,K1,K2,K3}.}
\end{figure}

Fig. 8 shows the benchmark plot comparing the 2D electron gas densities obtained in this work with the as-grown single-2D electron gases reported in the literature for N-polar III-nitride semiconductor heterostructures on various substrates: SiC, Sapphire, GaN and AlN. The GaN/AlGaN 2D electron gases fabricated in this work exhibit the highest densities among all N-polar III-nitride heterostructures: they are more than three times larger than highest 2D electron gases observed so far in N-polar III-nitride heterostructures grown on AlN substrates and only an order of magnitude below the intrinsic crystal limit of $\approx10^{15}$cm$^{-2}$. Furthermore, in the case of 4 period SL, $R_H$ reach 269 Ohm/sq which is one of the lowest reported to date for N-polar 2D electron gas channel. 

In the context of the device applications, the high-density 2D electron gas can be highly desired\cite{Mi}. In particular, the high-density 2DEG reduces the parasitic series resistances in the source/drain region which leads to an increasing of the on-currents, transconductance and, in consequence, the power output. For example, K.-S. Im et al.\cite{KS} used the extremely high-density 2DEG ($>10^{14}$cm$^{-2}$) as a source and drain in the Ga-polar GaN metal-oxide-semiconductor field-effect transistor (MOSFET) (which was based on the AlGaN/GaN heterostructure). Due to this, the authors were able to realize a high drain current and extrinsic transconductance. In addition, a high-density 2DEG may reduce the effect of the surface and buffer traps\cite{Ma1,Ma2}. However, there can be one drawback of the developed SL structure, namely the thermal resistance of the device may increase after introduction of the SL structure. This is because the actual SL structure likely includes high Al content AlGaN (see Fig. 6(a)) which exhibits low thermal conductivity. Nevertheless, we expect that due to very high 2DEG the on-resistance ($R_{on}$) will be much lower than in conventional devices. Due to lower $R_{on}$, the operation temperature of the device should be also lower because of the suppression of Joule heat generation ($Q=I^2R_{on}t$, where $I$ is the current and $t$ is the time) under constant current mode operation. Beside the transistor applications, high-density 2DEG systems are also attractive for infrared plasmonic applications\cite{S1,S2} or “charge gain” devices\cite{S3,S4}. In particular, there are many reports about the application of AlGaN/GaN heterostructures in terahertz plasmonic sources and detectors\cite{Ah,Spi,Ba,Bo,Dy,Ci,Sh}. It is obvious that the extremely high-density 2DEG can significantly boots the performance of these devices (since the frequency of 2D plasmons is proportional to 2D electron concentration). Thus, the ability of realization of N-polar III-nitride heterostructures with very high-density 2DEG may be desirable for various device applications.

In summary, we reported on the observation of the extremely high-density 2D electron gas in N-polar AlGaN/GaN heterostructures grown on sapphire substrates. We found that the introduction of the GaN/AlN superlattice back barrier between the GaN buffer layer and AlGaN barrier layer leads to a giant enhancement of 2DEG density at the GaN/AlGaN interface from $3\times10^{13}$cm$^{-2}$ (without SL) to $1.4\times10^{14}$cm$^{-2}$. The increase of 2DEG density correlated well with the changes of the wafer warp parameter, which suggested that the strains are responsible for the enhancement of 2DEG density. More precisely, SL increases compressive stress which leads to a reduction of tensile strain and piezoelectric polarization in the GaN channel layer. However, the changes in piezoelectric polarization of GaN channel layer solely are probably insufficient to fully explain the observed extremely high density of 2DEG.
Simultaneously, the room temperature electron mobility (at the 2D electron gas densities $1.4\times10^{14}$cm$^{-2}$) was 169 cm$^2$/Vs which led to the low Hall sheet resistance of 264 Ohm/sq (one of the lowest reported so far for N-polar 2D electron gas channel).

\begin{acknowledgments}
The authors expresses gratitude to Norikazu Nakamura for his kind support and discussions The authors would like to thank Yoshiharu Kinoue and Takaaki Sakuyama for their support in experiments. 
\end{acknowledgments}


\end{document}